# Evac-Cast: An Interpretable Machine-learning Framework for Evacuation Forecasts Across Hurricanes and Wildfires

Bo Li[a*], Chenyue Liu[a], Ali Mostafavi[a]

[a] UrbanResilience.AI Lab, Zachry Department of Civil and Environmental Engineering, Texas A&M University, College Station, TX, 77843, USA

[*] Corresponding author: Bo Li, libo@tamu.edu, UrbanResilience.AI Lab, Zachry Department of Civil and Environmental Engineering, Texas A&M University, College Station, TX, 77843, USA

**Abstract**

Evacuation is critical for disaster safety, yet agencies lack timely, accurate and transparent tools for evacuation prediction. This study introduces Evac-Cast, an interpretable machine learning framework that predicts tract-level evacuation rates using over 20 features derived from four dimensions: hazard intensity, community vulnerability, evacuation readiness, and built environment. Using XGBoost model trained on multi-source, large-scale datasets for two hurricanes (Ian 2022, Milton 2024) and two wildfires (Kincade 2019, Palisades–Eaton 2025), Evac-Cast achieves mean absolute errors of 4.5% and 3.5% for hurricane and wildfire events, respectively. SHAP analysis reveals a consistent feature importance hierarchy across hazards, led by hazard intensity. Notably, the models perform well without explicit psychosocial variables, suggesting that macro-level proxies effectively encode behavioral signals traditionally captured through time-consuming surveys. This work offers a survey-free, high-resolution approach for predicting and understanding evacuation in hazard events, which could serve as a data-driven tool to support decision-making in emergency management.

1. Introduction

Natural hazards are escalating in both frequency and severity, leading to rising loss of life and property damages. In recent years, natural disasters causing damage amounting to billions of dollars and inflicting loss of human life have occurred with increasing frequency in the United States, placing higher demands on emergency management (National Centers for Environmental Information, 2025). Evacuation is the primary protective action that transforms early warnings into saved lives. In the face of impending hazards, timely relocation of individuals from high-risk areas increases the availability of transportation infrastructure for incoming relief operations, reduces congestion across communication channels essential for emergency coordination, and allows first responders to focus efforts on assisting individuals who are unable to evacuate independently. Accurate, timely, and fine-grained forecasts of evacuation rates lie at the nexus of preparedness, real-time response, and post-impact search-and-rescue (SAR) efforts. With reliable evacuation estimates, agencies can pre-stage resources such as transportation and fuel, optimize the timing of traffic management strategies including contra-flow, and properly scale shelter capacity. Critically, evacuation prediction can also inform the assignment of SAR assets to the neighborhoods anticipated to exhibit low departure rates, thereby translating prediction into targeted life-saving interventions. Beyond forecasting, identifying the underlying factors that drive evacuation decisions is essential for crafting more effective public messaging and addressing systemic barriers. Accordingly, the study of evacuation decision-making becomes a strategic imperative for enhancing disaster response and saving lives.

Despite foundational understanding built from decades of research, accurate evacuation prediction must still overcome several key challenges. First, the conventional understanding of evacuation behaviors is informed primarily by psychological and social attributes, emphasizing roles such as risk perception, social ties, as well as prior experience (Buylova, Chen, Cramer, Wang, & Cox, 2020; Lovreglio, Ronchi, & Nilsson, 2016; Metaxa-Kakavouli, Maas, & Aldrich, 2018; Riad, Norris, & Ruback, 1999; Thompson, Garfin, & Silver, 2017). While these studies provide nuanced explanations of the mechanisms translating such factors into evacuation decisions, the narrow focus can lead to an incomplete understanding of the full spectrum of drivers. Moreover, these attributes are difficult to collect in time and at a large scale, which makes them less applicable for the rapid, operational forecasting needed before an approaching hazard. Other dimensions, such as socio-demographics and hazard-related conditions, have also been incorporated into consideration, yet factors from those dimensions are still not fully explored or are examined separately, lacking an integrated framework to assess their collective predictive power (Tanim, Wiernik, Reader, & Hu, 2022; Thompson et al., 2017). Second, traditional methods to predict and understand hazard evacuation behaviors frequently rely on statistical models that assume linear relationships (e.g.(Huang, Lindell, & Prater, 2017; Huang, Lindell, Prater, Wu, & Siebeneck, 2012; Thakur, Ranjitkar, & Rashidi, 2022a)), while recent studies have demonstrated that evacuation decision are influenced by complex, non-linear interactions that exist among the multi-faceted features (Sun, Huang, & Zhao, 2024; Zhao, Lovreglio, & Nilsson, 2020). Third, the scope of current evacuation analysis usually focuses on a single hazard type, which limits the generalizability of findings and prevents the identification of universal patterns that may transcend specific disaster scenarios.

To address these identified gaps, this study introduces a novel framework to advance the prediction and understanding of evacuation patterns across multiple hazard types. We move beyond the limitations of prior studies by employing an interpretable machine learning approach capable of capturing the complexity embedded in the diverse factors that may influence evacuation decisions. Our comprehensive framework, Evac-Cast, integrates readily accessible data from four key dimensions: hazard intensity, community vulnerability, evacuation readiness, and the built environment. Applying this framework to four major hazard events (two wildfires and two hurricanes) using a gradient-boosting model coupled with interpretability techniques, this study aims to (1) develop a high-accuracy predictive model for hazard evacuation that does not rely on traditional, difficult-to-acquire behavioral attributes, (2) leverage model interpretability to identify and rank the most significant features driving evacuation behavior, (3) assess the generalizability of evacuation drivers through cross-hazard comparison to identify universal response patterns.

The findings demonstrate that high predictive accuracy can be achieved without the inclusion of explicit micro-level behavioral variables. This outcome challenges the conventionally emphasized role of those factors in evacuation prediction, suggesting their influence may be less direct than previously understood. This research thus opens new avenues for more streamlined, scalable, and effective evacuation forecasting, and it contributes a new, data-driven perspective on the universal drivers of human behavior in the face of disaster. The model and findings presented in this paper advance evacuation research on four fronts. First, by deploying gradient-boosted trees on rich and readily available hazard, infrastructure, and socio-demographic data, the model captures the non-linear interactions that conventional linear or survey-based approaches miss, boosting predictive power. Second, integrating the model with SHAP (SHapley Additive exPlanations) enhances interpretability by identifying the key factors influencing evacuation rates, thereby providing actionable insights for emergency managers on where and how to intervene to expedite population movement during emergencies. Third, training and testing Evac-Cast on both hurricanes and wildfires reveals a stable hierarchy of drivers across hazards, demonstrating that the tool generalizes across regions and threat types rather than being tuned to a single scenario. Finally, the model attains high out-of-sample accuracy (MAE ≈ 3.5%–4.5%) without explicit psychosocial variables, challenging decades of evacuation literature that foregrounds micro-level behavioral factors and showing that much of their influence is already encoded in readily measurable contextual features. Together, these contributions position Evac-Cast as a scalable, transparent, and transferable decision-support asset for real-time evacuation planning.

The remainder of this paper is organized as follows: Section 2 provides a review of state-of-the-art literature, focusing on evacuation predictors, as well as methods and data to study evacuation behaviors. Based on the discussion, this section outlines the key gaps of current research. Section 3 first proposes the framework of the evacuation prediction model, then provides a brief introduction on the studied events, data, and methodology applied in the study. Section 4 presents result of the proposed model on hurricane and wildfire events. Section 5 discusses primary findings and summarizes contributions and limitations of the study.

## 2. Literature review

This section provides a comprehensive literature review on the factors influencing hazard evacuation decisions and the employed data and methods. It highlights the evolution of research in this field, identifies critical gaps in current knowledge, and establishes the foundation for the framework proposed in this paper.

2.1 Predictors of hazard evacuation decisions

The decision to evacuate before a hazard strikes is a complex behavioral process influenced by a multitude of interacting factors. Research has explored these factors from various perspectives to build a primary understanding. Early and ongoing research has significantly emphasized the role of individual psychological attributes and social contexts. Key psychological factors include risk perception, self-efficacy and response efficacy (Dash & Gladwin, 2007; Demuth, Morss, Lazo, & Trumbo, 2016; Forrister et al., 2024; Morss, Cuite, & Demuth, 2024; Thakur, Ranjitkar, & Rashidi, 2022b). Meta-analyses show that higher risk perception and official warnings are strong predictors of evacuation (Tanim et al., 2022). Social factors such as social ties, social cues (observing the evacuation behavior of neighbors and family), and social norms also play a crucial role (Lovreglio et al., 2016; Roy, Hasan, Abdul-Aziz, & Mozumder, 2022). Theoretical frameworks like the Protective Action Decision Model (PADM) established by Lindell and Perry (1992) and Lindell and Perry (2003) have been widely used to explain how these variables influence evacuation decision-making through a set of psychological processes (e.g.(X. A. Zhang & Borden, 2024)).

Also, a significant body of literature demonstrates that socio-demographic characteristics are critical determinants of evacuation behavior (Rambha, Nozick, & Davidson, 2021). Variables such as age, gender, race/ethnicity, level of education, presence of children, and home ownership consistently emerge as influential (Collins, Polen, McSweeney, Colón-Burgos, & Jernigan, 2021; Guan & Chen, 2021; Halim, Jiang, Meng, Mozumder, & Yao, 2024; Wong, Pel, Shaheen, & Chorus, 2020). For example, communities with higher non-white rate tend to have lower evacuation rate (Sun, Forrister, et al., 2024), while higher education level are often associated with a higher likelihood of evacuation (Kuligowski et al., 2022). Socio-economic status, such as income levels and homeownership, also significantly impacts evacuation decisions (Rashid, Tirtha, Eluru, &

Hasan, 2025). Higher income can facilitate access to resources like transportation and temporary lodging, whereas homeownership is often negatively correlated with evacuation, potentially due to greater place attachment or property concerns (Hasan, Mesa-Arango, Ukkusuri, & Murray-Tuite, 2012). Vehicle access usually facilitates evacuation likelihood by removing transportation barriers for households (Anyidoho, Davidson, Rambha, & Nozick, 2022; Golshani, Shabanpour, Mohammadian, Auld, & Ley, 2019).These factors often contribute to a household's overall social vulnerability and reflect evacuation readiness.

Hazard-related characteristics is another key indicator of evacuation decision. Key aspects include hazard intensity, proximity to hazards, speed of onset, expected duration, and spatial extent (Anyidoho et al., 2022; León, Gubler, & Ogueda, 2022; Lindell, Lu, & Prater, 2005; Mozumder & Vásquez, 2018; Wu et al., 2022). For example, R. Zhang, Liu, Xu, Xu, and Chen (2024) studied factors influencing evacuation decisions under different flash flood characteristics. The communication of hazard information and protective action recommendations is also a vital factor. Official warnings and evacuation orders (mandatory versus voluntary) are strong predictors of compliance (Huang, Lindell, & Prater, 2016; Sadri, Ukkusuri, & Gladwin, 2017; Younes, Darzi, & Zhang, 2021). Built environment features also impact the feasibility and appropriateness of evacuation, which further influences residents' choices to evacuate or to shelter in place (Bakhshian & Martinez-Pastor, 2023). Studies show that built environment features, such as population density (Golshani et al., 2019), land parcel size (Wu et al., 2022), housing construction type and transportation characteristics can significantly alter evacuation possibilities (Anyidoho et al., 2022; Fusco, Zhu, & Yang, 2025).

2.2 Applied methods and data in hazard evacuation research: challenges and opportunities

2.2.1 Evacuation analysis data

Traditionally, research has relied heavily on post-event surveys, interviews, and traffic count data to gather data on evacuation behaviors, decision-making processes, and influencing factors. Information on household evacuation decisions and actions has frequently been gathered through surveys, such as questionnaires, interviews, and panel surveys (Kuligowski et al., 2022; Lindell et al., 2005; Lindell, Murray-Tuite, Wolshon, & Baker, 2018; Murray-Tuite, Yin, Ukkusuri, & Gladwin, 2012). For example, Lindell et al. (2005) utilized mail survey to gather extensive information on household evacuation decisions after Hurricane Lili. Similarly, telephone survey collected helped model individual-level evacuation decisions from Hurricane Sandy (Sadri et al., 2017). Kuligowski et al. (2022) collected online survey data from households impacted by the 2019 Kincade fire in California about their evacuation decisions and potential influencing factors. Beyond survey data, research also applied traffic detector data to study evacuation behaviors in hurricanes and wildfire scenarios. For example, Woo, Hui, Ren, Gan, and Kim (2017) characterized the 2016 Fort McMurray wildfire evacuation using traffic count and flight data. J. Li and Ozbay (2015) counted real-world traffic data to analyze evacuation traffic patterns during Hurricane Irene in New Jersey.

Surveys and interviews, while valuable for capturing individual perceptions and decision-making processes, often suffer from small sample sizes, recall bias, and delays in data collection, as they are typically conducted after an event. This limits their ability to provide real-time insights into evacuation dynamics. Although traffic count data offer objective measures of vehicle movements, it is usually restricted to major highways and lacks demographic context, leading to an incomplete picture of evacuation patterns.

Emerging technologies are offering new opportunities to overcome the limitations of traditional evacuation research. One promising avenue is the application of location intelligence data derived from anonymized mobile devices and GPS records. Location-based data provides large-scale, high-resolution insights on population movements in a timely manner, which enables human mobility tracking during hazard events. For example, Yabe, Sekimoto, Tsubouchi, and Ikemoto (2019) used GPS data from more than 1 million mobile phone in four earthquake events to characterize evacuation patterns. Wu et al. (2022), Washington, Guikema, Mondisa, and Misra (2024), and Liu, Zhang, Jiang, and Zhao (2025) proposed and improved algorithms to model evacuation characteristics, such as evacuation rate and evacuee types, using large-scale GPS data. B. Li and Mostafavi (2022) synchronized hurricane preparedness and evacuation based on location intelligence data to identify the hotspots of vulnerability during Hurricane Harvey. Lee, Chou, and Mostafavi (2022) characterized evacuation return and home-switch stability as milestones of disaster recovery based on location-based data. Ultimately, location intelligence data unlocks the ability to provide more accurate and scalable insights into evacuation dynamics.

2.2.2 Evacuation prediction methods

To understand and predict evacuation behaviors during hazard events, regression models are the most commonly used approach in various scenarios, such as hurricanes (Lindell et al., 2005; Morss et al., 2024), earthquakes (Ao, Huang, Wang, Wang, & Martek, 2020) and wildfires (Kuligowski et al., 2022; Wu et al., 2022). However, those models often rely on the linear relationships between input features and evacuation decisions, limiting their ability to capture the complex, non-linear dynamics of human behavior in crisis situations. To address this limitation and enhance predictive performance, some studies have explored more advanced machine learning techniques, such as support vector machines, tree-based algorithms, and neural networks to better handle nonlinearity (Balboa, Cuesta, González-Villa, Ortiz, & Alvear, 2024; Ebrahimi, Sattari, Lefsrud, & Macciotta, 2023; Xu et al., 2023). While these models offer improved predictions, they often operate as "black boxes", making it difficult to interpret the underlying decision-making processes. Therefore, interpretable machine learning has emerged as a promising approach. For example, Sun, Huang, et al. (2024) developed an enhanced logistic regression model for household hurricane-evacuation decisions by linking low-depth decision trees to detect nonlinear thresholds. Zhao et al. (2020) used a random forest to model individuals' pre-evacuation choices in a fire drill, then applied variable importance ranking and partial dependence plots to reveal key factors. Strong nonlinear effects and interactions have been revealed from those studies, highlighting the necessity of using interpretable machine learning methods to improve predictive accuracy while maintaining model transparency.

2.3 Point of departure

Despite valuable insight gained from decades of research, evacuation behavior research remains constrained by several methodological and empirical limitations that hinder their applicability in operational contexts. Much of the literature relies on post-event surveys to capture psychosocial variables that are costly to collect, prone to recall bias, and unavailable during the crucial pre-impact window. Moreover, analysis typically impose linear statistical models that fail to capture the non-linear interactions suggested by theory. Even recent applications of complex machine learning models, while improving predictive accuracy, often treat the model as a black box, offering insufficient transparency to guide emergency planning. Additionally, existing studies frequently focus on a narrow subset of predictors while overlooking the broader, interconnected factors that shape evacuation decisions. Further, many existing models are developed in the context of a single hazard type, which limits their generalizability and hinders the discovery of cross-cutting patterns in evacuation behavior. Together, these limitations inhibit both scientific synthesis and the operational translation of findings. Fortunately, the increasing availability of high-resolution mobility, hazard information, census and infrastructure datasets presents new opportunities to overcome these challenges. Building on these developments, this study introduces an interpretable machine learning framework that integrates diverse data sources to reveal the multi-dimensional drivers of evacuation behavior, while enhancing both the transparency and reliability of predictive modeling for practical use in disaster response.

## 3. Data and method

3.1 Research flow

As shown in Fig.1, this study proposes and implements a comprehensive framework for predicting hazard evacuation rates and uncovering their underlying drivers using interpretable machine learning techniques. Applying multi-sourced large datasets, we incorporate a broad set of features representing four critical dimensions: community vulnerability, evacuation readiness, hazard intensity, and built environment characteristics. These multifaceted inputs are used to train an interpretable machine learning model (XGBoost combined with SHAP) to predict evacuation rates for two distinct hazard types. This approach enables both accurate evacuation forecasts at a fine spatial resolution and the identification of key factors influencing evacuation through model interpretability. In addition, we evaluate the generalizability and transferability of the identified patterns across different hazard types, contributing to a more robust and scalable understanding of evacuation dynamics.

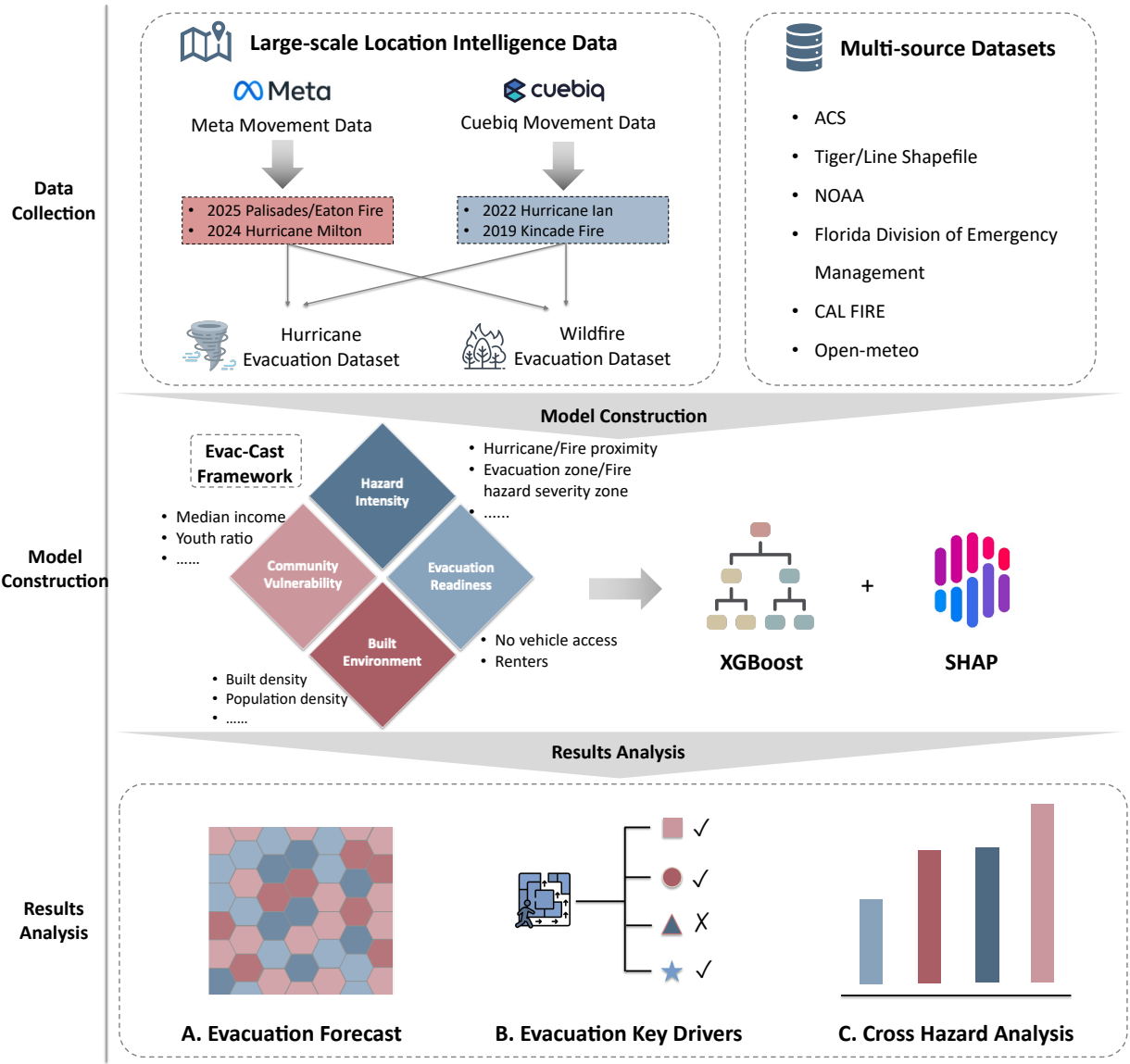

**Fig.1** Research flow for establishing Evac-Cast model to predict and interpret hazard evacuation using large-scale datasets and interpretable machine learning techniques.

3.2 Research context

To ensure a robust and generalizable analysis, this study focuses on four large-scale natural hazards events: two hurricanes (i.e., Ian and Milton) in Florida and two wildfires (i.e., Kincade and Palisades–Eaton fires) in California. These states were selected due to high frequency of natural hazards and vulnerability to these respective hazards, providing a rich context for examining evacuation behavior.

The state of Florida's extensive coastline and low-lying geography make it highly prone to hurricanes, which has repeatedly result in significant damages and disruption (Florida Climate Center). This study examines two major events: Hurricane Ian in 2022 and Hurricane Milton in 2024. Hurricane Ian made landfall in southwestern Florida as a Category 4 storm, triggering evacuation orders or warnings for more than 2.5 million residents (Brooks, 2022; National Hurricane Center, 2023). More recently, Hurricane Milton made landfall as a powerful Category 3 storm near Siesta Key, and its rapid intensification and projected path triggered widespread and urgent evacuation orders for millions of residents in a densely populated region of Southwest Florida (Julia Reinstein, 2024; National Weather Service, 2024).

California's climate and topography create a significant threat of destructive wildfires, and the risk is intensified by the growth of housing at the wildland–urban interface (Manzhu Yu, 2025). This study examines two significant events: the Kincade Fire in 2019 and the devastating January 2025 Southern California wildfires. The Kincade Fire in Sonoma County burned more than 77,000 acres, triggering the largest evacuation in the county's history (County of Sonoma, 2020). The January 2025 event involved multiple fires, with the two largest being the Palisades Fire and the Eaton Fire in Los Angeles County. Burning simultaneously, the Palisades Fire devastated coastal communities while the Eaton Fire swept through foothill neighborhoods. Together, these two fires forced a vast scale of evacuation and destroyed thousands of structures, representing a complex, multi-front urban evacuation scenario (Charlotte Phillipp, 2025).

3.3. Evacuation data

This study uses large-scale, anonymized mobility data to objectively measure evacuation behavior. This approach overcomes the limitations of traditional survey methods, such as recall bias and the intention–behavior gap, by capturing population movements at a high temporal and spatial resolution. We use evacuation rate as the primary response variable for the Evac-Cast model. To define the spatial scope of the analysis, this study incorporates only census tracts located within counties that issued an official evacuation order for each respective hazard event. The definition of the evacuation period was tailored to the specific nature of each hazard type. For hurricanes, which have a forecast period, the evacuation window is defined as the three days leading up to landfall. For wildfires, which often occur with little to no warning, we define the evacuation period as beginning on the date when the fire was first reported, and take the peak evacuation rate recorded during that period as our proxy measure. For the earlier events, Hurricane Ian (2022) and the Kincade Fire (2019), this study utilizes location intelligence data from Cuebiq. This dataset consists of anonymized location signals derived from mobile devices, providing a broad view of population presence and movement patterns before and during the hazard events. Evacuation rates at the census tract level are estimated from anonymized Cuebiq mobility data by identifying significant shifts in users' primary evening locations. A stable home location for each device was established with the Resilitix AI home-detection method, which clusters recurrent weekday-evening stay points observed during the five weeks preceding the hazard and cross-validates cluster stability, permitting up to a two-week tolerance for missing or inconsistent observations. Building on additional Resilitix AI mobility inference rules, evacuation status was evaluated with a five-day rolling window in which each user's primary location was reassessed daily. A device is deemed "evacuated" when it remains outside its pre-event home census tract for at least three consecutive days, and "returned" once presence at that same census tract is re-established for three or more consecutive days. These individual-level trajectories were then aggregated to derive daily evacuation rates across the studied region.

For the more recent events, the January 2025 Wildfires (Palisades–Eaton) and Hurricane Milton (2024), we use Facebook's 'Population During Crisis' data to calculate evacuation rate. The dataset reports the number of Facebook users observed in Bing tile level 14 (roughly equivalent to the size of 6 × 6 city blocks) during the hazard period and the baseline period. We aggregate each tile to a census tract using its centroid coordinates. Since user counts are provided in 8-hour intervals, we average the data records to create a single daily value. Then we calculate the daily evacuation rate for each tract based on data during the hazard and baseline periods. Fig. 2 displays the spatial distribution of evacuation rates for the four events of interest, overlaid with corresponding hurricane paths or fire perimeters to illustrate the spatial range of hazards.

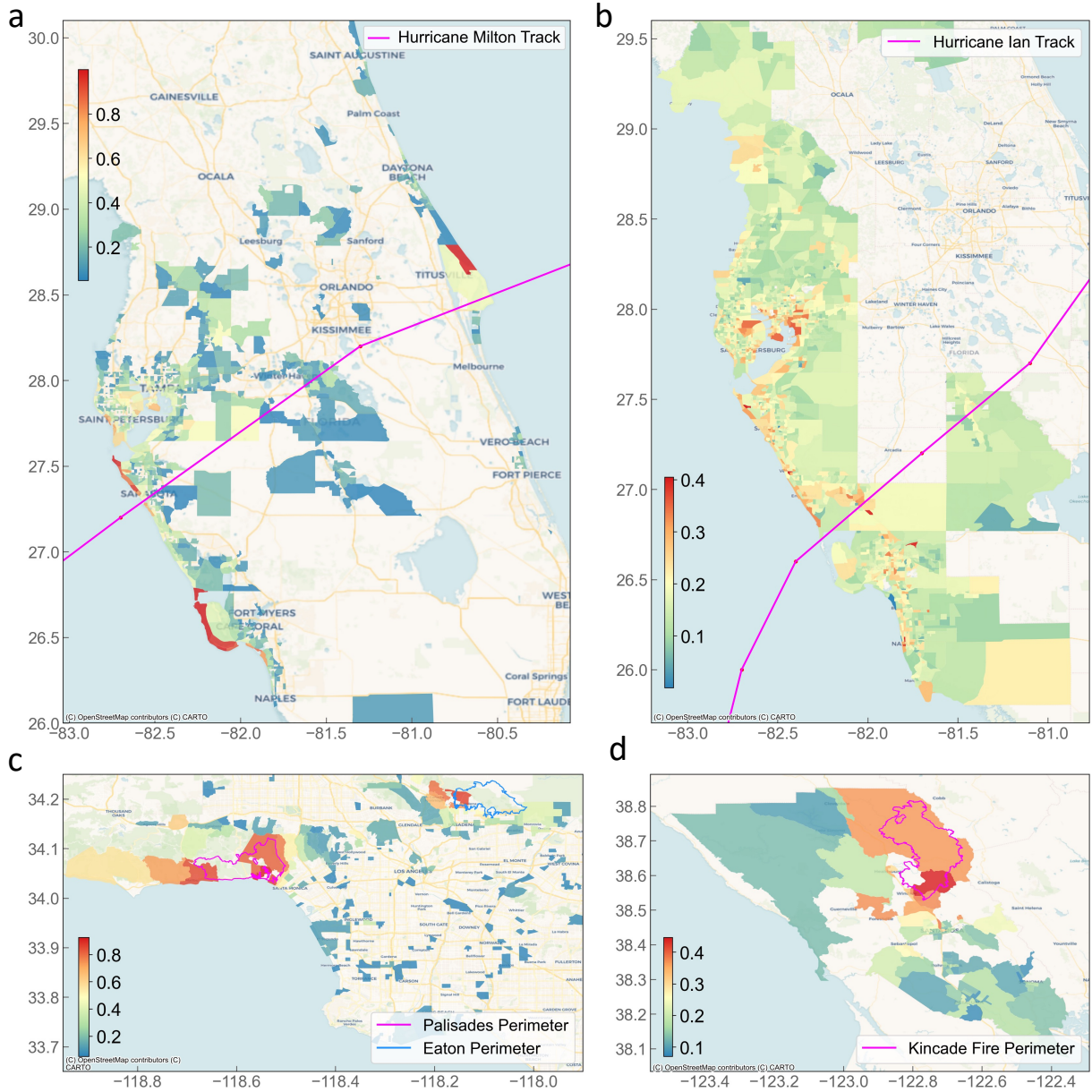

**Fig.2** Spatial distribution of evacuation rate during a) Hurricane Milton, b) Hurricane Ian, c) the Palisades–Eaton Fire, and d) the Kincade Fire. Hurricane paths and fire perimeters are overlaid to depict spatial range of hazards.

3.4 Evacuation predictors

To build a comprehensive predictive model, this study incorporates a wide range of variables that have been identified in previous literature as potential drivers of evacuation behavior. We categorize these variables into four distinct dimensions: hazard intensity, community vulnerability, evacuation readiness, and the built environment (Fig.3). The features are derived from multiple authoritative sources, including the American Community Survey (ACS), National Oceanic and Atmospheric Administration (NOAA), and California Department of Forestry and fire Protection (CAL FIRE). A brief description of variables and their sources is provided in Table 1. All variables can be available in near-real time from public or commercial feeds, satisfying the timeliness constraint of operational forecasting, and are defined consistently across hazards, enabling the cross-event generalizability.

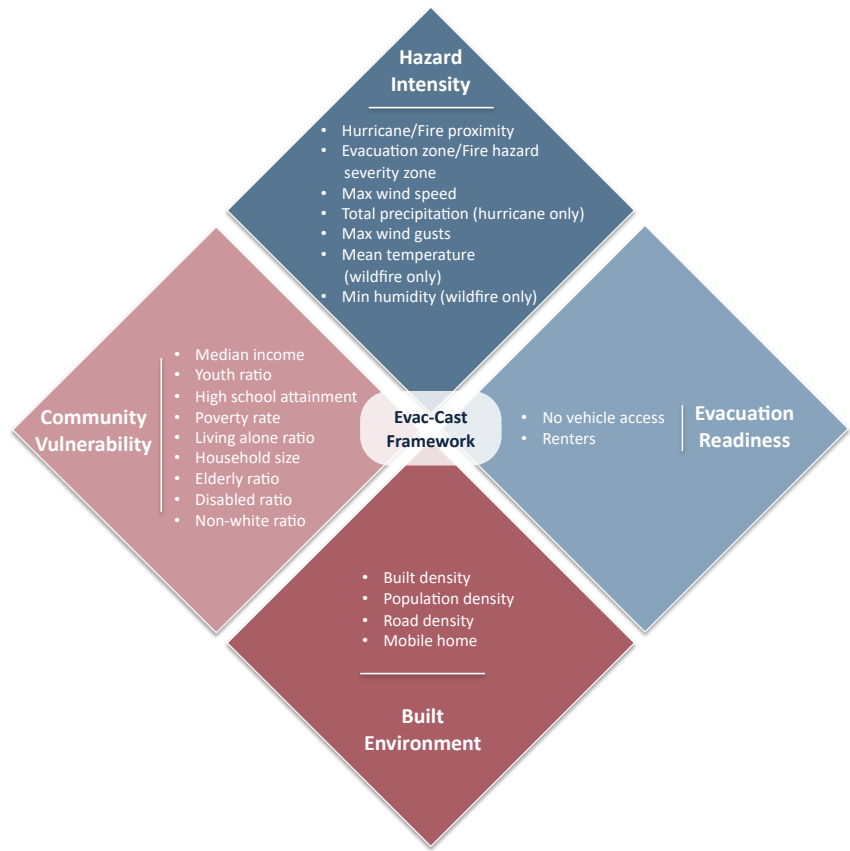

**Fig.3** Evac-Cast framework for evacuation prediction in natural hazard scenarios. The framework contains four dimensions: hazard intensity, community vulnerability, evacuation readiness and built environment. Each dimension encompasses a set of relevant features that collectively capture the physical threat, sociodemographic context, logistical capacity, and physical built facilities influencing evacuation extent.

Table 1. Definitions and sources of variables

| Variable | Description | Source |
|---|---|---|
| **Built environment** | | |
| Building density | Ratio of total building footprint area to the total land area of a census tract. | Bing Maps building footprints |
| Population density | Number of people per square kilometer within a census tract. | ACS, TIGER/Line® Shapefiles |
| Road density | Total length of primary and secondary roads (in km) per square kilometer of a census tract. | TIGER/Line® Shapefiles |
| Mobile home | Percentage of housing units that are mobile homes. | ACS |
| **Hazard intensity** | | |
| Hurricane proximity | Minimum distance (in km) from a census tract to the hurricane's path. | NOAA |
| Evacuation zone | Official evacuation zone designation for a census tract. | Florida Division of Emergency Management |
| Max wind speed | Maximum sustained wind speed (at 10 m height) recorded for a census tract. | Open-meteo |
| Total precipitation | Sum of precipitation recorded for a census tract. | Open-meteo |
| Max wind gusts | Maximum wind gust speed (at 10 m height) recorded for a census tract. | Open-meteo |
| Fire proximity | Minimum distance (in km) from a census tract to the fire perimeter. | CAL FIRE |
| Mean temperature | Mean temperature, used for wildfire events. | Open-meteo |
| Min humidity | Minimum daily relative humidity, used for wildfire events. | Open-meteo |
| Fire hazard severity zone | Official CAL FIRE designation of wildfire hazard severity for a census tract. | CAL FIRE |
| **Community vulnerability** | | |
| Median income | Median household income of a census tract. | ACS |
| Youth ratio | Percentage of the population under the age of 18. | ACS |
| High school attainment | Percentage of the population, 25 years and over, with at least a high school diploma. | ACS |

| Poverty rate | Percentage of the population living below the poverty level. | ACS |
| --- | --- | --- |
| Living alone ratio | Percentage of households occupied by a single person. | ACS |
| Household size | The average number of people per household. | ACS |
| Elderly ratio | Percentage of the population aged 65 and over. | ACS |
| Disabled ratio | Percentage of the civilian noninstitutionalized population with a disability. | ACS |
| Non-white ratio | Percentage of the population identifying as a race other than white alone. | ACS |
| **Evacuation Readiness** | | |
| Renters | Percentage of occupied housing units that are renter-occupied. | ACS |
| No vehicle access | Percentage of households with no available vehicles. | ACS |
| **Response Variable** | | |
| Evacuation rate | Percentage of the population that evacuated from a given census tract during the hazard event. | Facebook movement data/Cuebiq |

### 3.4.1 Hazard intensity

This dimension quantifies the direct physical threat posed by natural hazard events. It includes dynamic features that describe the magnitude and proximity of the hazard, which serve as primary signals of risk for residents in affected areas.

- Hurricane proximity: This feature measures the distance from a given census tract to the hurricane's path, providing a direct measure of exposure. We first retrieved hurricane track data from NOAA's historical hurricane tracks database (National Oceanic And Atmospheric Administration), which provides cyclone positions and intensities at regular intervals. Then we overlaid those track points against the study area's census tract locations, and computed the distance from every tract centroid to the hurricane center at each time step. The shortest distance between the hurricane center and each census tract serves as its hurricane proximity value.

- Evacuation zone: This feature represents evacuation priority, which classifies areas from Zone A (highest risk, evacuation first) to Zone E (lowest risk). We retrieved Florida's official statewide evacuation zone map from the Florida Division of Emergency Management (Florida Division of Emergency Management). Then we performed a spatial intersection analysis with census tract boundaries to assign an evacuation zone to each census tract. For tracts overlapping multiple zones, the zone covering the largest portion of the tract was selected.

- Fire hazard severity zone: This feature classifies land as moderate, high, and very high fire hazard severity zones based on the likelihood and potential intensity of a wildfire, as designated by CAL FIRE (CALFIRE, 2024). We performed a spatial join analysis to assign fire hazard severity classification to each census tract from the nearest fire hazard severity zone polygon.

- Weather conditions: We included a suite of meteorological features to provide a comprehensive profile of the hurricanes' characteristics and the ambient conditions influencing fire behaviors. Weather data was collected for each census tract using its centroid via the Open-Meteo API, a powerful tool that aggregates data from high-resolution models from national weather services (Open-Meteo). For hurricane events, we first collected daily observations of max wind speed, max wind gusts, and precipitation sum. Then we aggregated these to the evacuation period by (i) taking the maximum value observed from wind speed and wind gust, and (ii) summing precipitation across the entire period. For wildfire events, the collected feature included daily mean temperature, minimum humidity, maximum wind speed, and maximum wind gust, which are critical factors in fire spread. Starting on the ignition date and ending when the peak evacuation rate is reached, we aggregated these daily observations by taking (i) the highest sustained wind speed, (ii) the highest wind gust, (iii) the lowest relative humidity, and (iv) the average daily temperature over that window.

- Fire proximity: This variable measures the distance from each census tract to the fire perimeter, providing a direct measure of exposure to an advancing wildfire. Fire Perimeters for Kincade and Palisades/Eaton are obtained from CAL FIRE 's Fire and Resource Assessment Program (CALFIRE). We calculated the minimum distance from the boundary of each census tract to the nearest fire perimeter using spatial operations.

3.4.2 Community vulnerability

This dimension captures the socio-demographic characteristics of the population within a census tract that can influence their capacity to prepare for, respond to, and recover from a natural hazard. Specifically, we took following features into consideration:

- Median income and poverty rate: Features of the financial resources available to households, which can facilitate or constrain evacuation.

- Youth and elderly ratio (%): These demographic features represent populations that may have unique needs or risk perceptions.

- High school attainment (%): A feature of educational level, which can correlate with access to information and resources.

- Living alone ratio (%) and household size: Features describing the household structure, which can impact the logistics and social dynamics of an evacuation decision.

- Disabled population (%): A feature representing the proportion of residents who may face significant mobility or health-related challenges during an evacuation.

All those features are retrieved from American Community Survey (U.S. Census Bureau) at census tract level. Features in the form of percentages are calculated using the absolute number divided by total population in the census tract.

3.4.3 Built environment

This dimension describes the physical characteristics of the human-made landscape, including the density and layout of development, which directly affects exposure to hazards and the feasibility of evacuation.

- Built density: This feature measures the concentration of structures as the ratio of built-up area to the total area in each census tract. Higher built density can indicate greater potential for loss and more complex evacuation logistics. We first retrieved building footprint data from Bing Maps (Bing Maps).Then we aggregated building footprints within each census tract boundary using spatial join, and summed the total areas of all buildings. Built density value was computed by dividing this total built-up area by the total land area of the census tract, providing a standardized measure of how intensively developed each tract is.
- Population density: This feature is calculated by dividing the total population of each census tract by its geographic area, using data from ACS (U.S. Census Bureau) and TIGER/Line Shapefiles (U.S. Census Bureau). Population density offers a standardized measure of how many people reside per unit of area in each tract, reflecting the extent of urban development.
- Road density: This feature is calculated by dividing the total length of roads by the area of each census tract, using data from TIGER/Line Shapefiles (U.S. Census Bureau). We intersected primary and secondary road networks with census tract boundaries, summed the length of road segments within each tract, and then divided the sum by the tract's total area. Road density serves as a proxy for the capacity of the transportation network to handle evacuating traffic.

- Mobile home (%): The percentage of housing units that are mobile homes. Residents of these structures are often considered more vulnerable to high winds and fire, which may increase their propensity to evacuate. The data was obtained from ACS (U.S. Census Bureau).

3.4.4 Evacuation readiness

This dimension reflects the underlying preparedness and logistical capacity of a community. While closely related to community vulnerability, these factors aim to capture key structural and resource-based factors that directly facilitate or impede a household's willingness and ability to evacuate. All features in this dimension are retrieved and calculated based on ACS.

- Renters (%): The proportion of households that are renters, which is a measure of home ownership. Tenure status can influence evacuation decisions, as renters may have different levels of attachment to a property and varying degrees of autonomy compared to homeowners.

- No vehicle access (%): The percentage of households without access to a personal vehicle. This is a critical logistical barrier that can severely constrain a household's ability to evacuate independently.

3.5 Model training and analysis

3.5.1 XGBoost model

XGBoost (Extreme Gradient Boosting) is an ensemble learning method that builds a strong predictive model by combining multiple decision trees to form a strong learner(Chen & Guestrin, 2016). To predict evacuation rates and understand their underlying drivers, this study implemented XGBoost regression models for hurricane and wildfire events separately. This hazard-specific approach allows for more accurate, context-sensitive predictions while enabling direct comparison of evacuation drivers across different hazard contexts.

For each hazard type, we randomly split the dataset into training (80%) and testing (20%) sets, reserving the testing set for a robust evaluation of model performance on unseen data. To mitigate imbalance in evacuation rates and improve model performance, we applied Synthetic Minority Over-Sampling Technique for Regression with Gaussian Noise (SMOGN) technique, which adaptively generates synthetic samples for under-represented portions of the target distribution through interpolation or adding Gaussian noise (Branco, Torgo, & Ribeiro, 2017). To identify the optimal hyperparameters for each hazard context, we conducted a random search with ten-fold cross validation on training data. The search explored a predefined grid of key parameters, including maximum tree depth, learning rate, gamma, minimum child weight. The combination that minimized the mean squared error was selected as the optimal set. After tuning, we trained the model with optimal hyperparameters and evaluated its performance on the testing sets.

3.5.2 SHAP analysis

While XGBoost model could provide highly accurate predictions, their complexity makes it difficult to understand the underlying mechanisms behind their outputs. To address this black-box problem, we employed Shapley Additive explanation (SHAP) for model interpretation. SHAP explains how each feature contributes to a model's prediction, quantifying its specific impact based on game theory (Lundberg & Lee, 2017). We calculated SHAP values for every feature across their respective test sets for both hurricanes and wildfires. By aggregating these values, we gained insights on the features' contributions towards shaping evacuation behaviors.

4 **Result and discussion**

4.1 Model performance

This study proposes a framework that incorporates four dimensions of features, hazard intensity, community vulnerability, evacuation readiness, and built environment, to predict evacuation rates under various hazard scenarios. Under this framework, two XGBoost models were trained and tested using multi-source data for hurricane and wildfire events. Table 2 presents the performance for both models using four evacuation metrics: root mean squared error (RMSE), mean absolute error (MAE), mean squared error (MSE), and coefficient of determination ($R^2$). For hurricanes, the model achieved an RMSE of 0.0668 and an MAE of 0.0452, indicating that the predicted evacuation rates deviate from the actual values by approximately 4.5%–6.7% on average. The R² value of 0.8074 suggests that over 80% of the variance in evacuation rates is explained by the model. For wildfires, the RMSE of 0.0559 and MAE of 0.0351 reflect an average prediction error of 3.5%–5.6%, and the R² of 0.8336 indicates strong explanatory power. Overall, these results confirm that Evac-Cast, the machine learning framework which utilizes readily accessible data, can predict hazard evacuation rates with a high degree of accuracy and practical significance for both hurricane and wildfire events.

Table 2. Model performance

| Hazard event | MSE | RMSE | $R^2$ | MAE |
| --- | --- | --- | --- | --- |
| Hurricanes | 0.0045 | 0.0668 | 0.8074 | 0.0452 |
| Wildfires | 0.0031 | 0.0559 | 0.8336 | 0.0351 |

4.2 Feature importance in shaping evacuation behaviors

Beyond predictive accuracy, this study advances the analysis by applying SHAP to uncover the key factors that could affect evacuation rates. Using SHAP values, we evaluate both individual feature importance and broader thematic patterns that influence evacuation behaviors. Fig. 4 presents the SHAP summary plots for the hurricane evacuation model, providing insights on the magnitude and direction of features towards shaping evacuation rates. Fig 4a ranks the features by their mean absolute SHAP value, which indicate the overall contribution of each feature toward evacuation rate. Fig. 4b is a SHAP beeswarm plot that illustrates how each feature contributes to individual predictions across all samples in the hurricane evacuation model. Each point represents a SHAP value for one sample, with its horizontal position indicating whether the feature increases or decreases the predicted evacuation rate. The color reflects the feature's actual value, showing how different feature levels contribute to model output. Collectively, it reveals both the direction and variability of features' influence on evacuation rate. For hurricanes, evacuation behavior is primarily shaped by variables directly reflecting the physical characteristics of the storm. Maximum wind speed and hurricane proximity emerge as the most influential features that are positively associated with proactive evacuation, emphasizing that residents respond decisively to the severity and immediacy of the approaching storm. Total precipitation and peak wind gusts also confirm their positive contributions towards promoting evacuation rate. This finding highlights the vital importance of not only generating timely and accurate forecasts but also ensuring that this information is effectively communicated at the local level. When residents are made aware of a storm's severity and its proximity to their location, they are more likely to evacuate promptly. The evacuation zone designation feature also ranks among the most important evacuation predictors, suggesting that official guidance is effective in motivating proactive evacuation behavior. Well-designed and consistently enforced evacuation zones can serve as a powerful tool for risk reduction, especially when paired with hazard forecasts and community outreach strategies that address access barriers and promote trust in public messaging.

Beyond hazard-related features, the result also reveals that community vulnerability, evacuation readiness and built environment features could affect evacuation behavior. Features such as renter ratio, population density and mobile home ratio also rank just below the top hazard intensity predictors. For example, higher renter ratio in a census tract is positively associated with its evacuation rate, which might be due to greater mobility or less attachment to property. The mobile home ratio also contributes positively, suggesting that residents in more vulnerable housing types are more likely to evacuate due to structural concerns. However, higher population density negatively contributes to evacuation rates, suggesting that densely populated areas are more likely to exhibit lower level of evacuation. A possible explanation is that high population density often corresponds to greater levels of urban development, where infrastructure is more robust and equipped with more shelter facilities within the area, so that shelter-in-place options are more readily available. Median income and living alone ratio also influence evacuation behavior. Higher median income is generally associated with increased predicted evacuation rates, likely reflecting greater access to resources, such as such as transportation, lodging, and emergency supplies that lower the barriers to evacuation. Additional contributing factors include living alone ratio, road density, youth population, as well as high school attainment ratio, while their influences are more moderate.

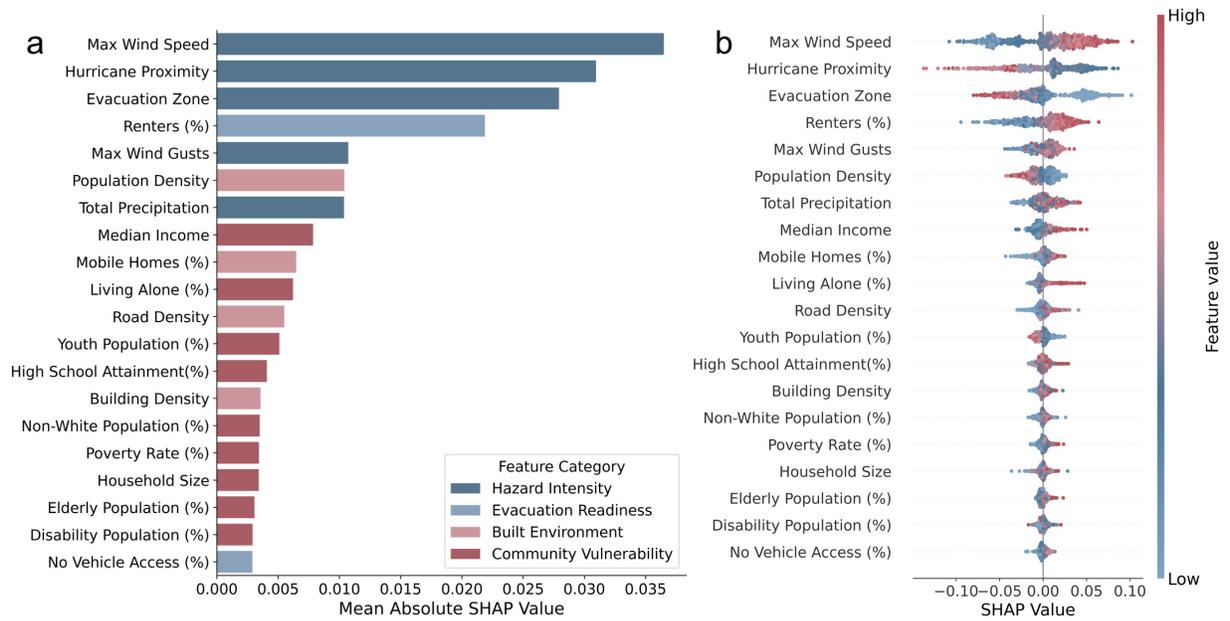

**Fig.4** SHAP summary plot for hurricane events. a) barplot ranks the features by their mean absolute SHAP values, showing the magnitude of feature importance. b) beeswarm displays how feature values affect individual predictions, showing the direction of feature contribution towards evacuation rate.

Fig. 5 presents the summary plot for wildfire evacuation model, offering insights into the relative importance and directional influence of various predictors. As shown in Fig. 5a, fire proximity, fire hazard severity zone and high school attainment ratio are the most influential feature that shaping evacuation behaviors. Fire proximity represents a direct measure of fire threats, capturing how physically close a community is to the wildfire. Smaller distances increase the evacuation rate, suggesting residents are more likely to evacuate when facing threats from acute, rapid escalating wildfires. Fire hazard severity zone reflects official classifications of wildfire risk based on comprehensive factors, such as vegetation, topography and historical fire patterns. Census tracts located within high-severity zones are associated with higher evacuation rates, suggesting that official classifications significantly influence residents' evacuation behaviors, which is similar to evacuation zone designation in hurricane events. Besides, other hazard intensity features, including mean temperature, minimum humidity, maximum wind speed and peak gust are also among the top influential features, indicating that hazard intensity serves an important role in promoting evacuation in wildfire events as well. Notably, high school educational attainment is the second-highest ranked feature in shaping wildfire evacuation behaviors. This suggests that communities with higher educational levels may have greater capacity to interpret warning messages, respond quickly, and access protective resources. In addition, median income and non-white ratio emerge as influential community vulnerability features, highlighting the role of socioeconomic barriers and disparities in shaping evacuation capacity.

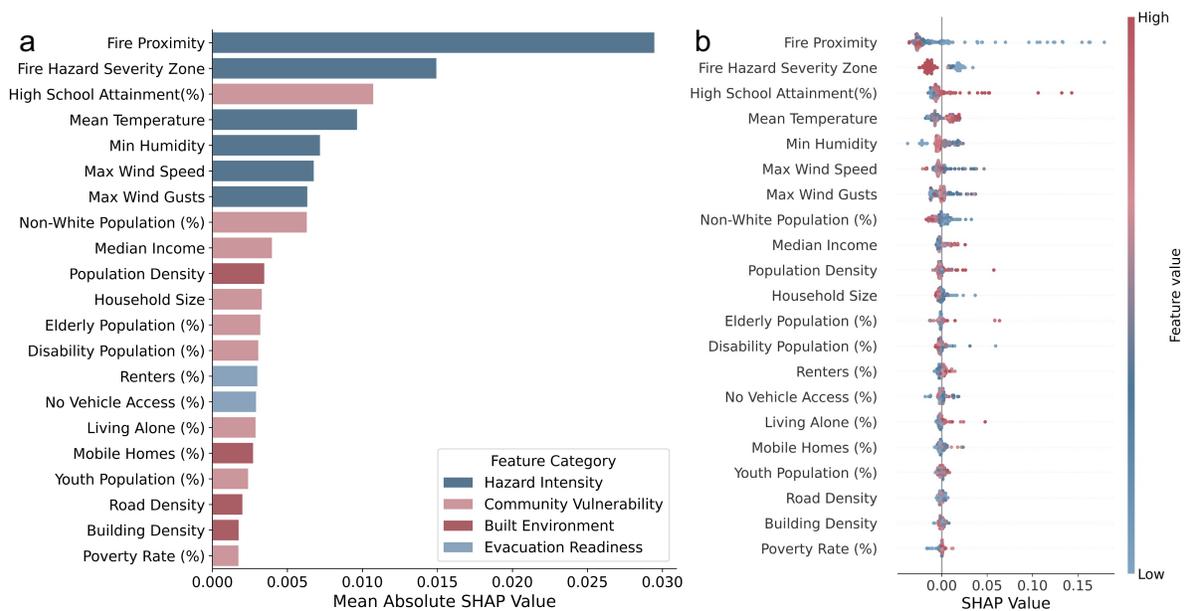

**Fig.5** SHAP summary plot for wildfire events. a) barplot ranks the features by their mean absolute values, showing the magnitude of feature importance. b) beeswarm displays how feature values affect individual predictions, showing the direction of feature contribution towards evacuation rate.

## 4.3 Cross-hazard analysis

To better understand the generalizability and context-specificity of evacuation behavior, we compared the relative contributions of feature dimensions across hurricane and wildfire events.

Fig. 6 shows the aggregated mean SHAP values for four dimensions, which enables high-level comparison of evacuation drivers across different hazard scenarios. The most significant finding from the cross-hazard analysis is the consistent ranking of the four feature dimensions across both hazard types. For both hurricane and wildfires, hazard intensity is the dominant driver of evacuation behavior, followed by community vulnerability, built environment and evacuation readiness. The consistent pattern suggests the existence of universal, structured driving factors for evacuation, regardless of the disaster's specific characteristics. This pattern indicates that communities exposed to more severe physical threats are most likely to exhibit higher evacuation rates. This finding reinforces a critical operational insight: evacuation behavior is primarily driven by the severity of the threat, highlighting the importance of timely, accurate, and clearly communicated risk indicators in prompting protective action. For emergency planners, this underscores the need to prioritize real-time hazard monitoring and dissemination as key levers for increasing evacuation compliance.

The second most influential dimension across both hurricane and wildfire events is community vulnerability, highlighting the important role of sociodemographic factors in shaping evacuation behavior. In both models, certain community vulnerability features, such as median income and high school attainment, are positively associated with evacuation rates. This suggest that sociodemographic status are important indicators for evacuation prediction, and people with greater social and economic resources are more likely to take proactive actions to evacuate. As a comparison, vulnerable communities, such as those with lower income or limited education level may take on low evacuation rates. These findings underscore the needs to closely monitor and address disparities in evacuation, and to ensure that emergency planning actively supports vulnerable groups through more accessible communication and equitable resource allocation.

Although the relative contribution of built environment and evacuation readiness is lower compared to hazard intensity and community vulnerability, these dimensions still play a meaning role in shaping evacuation. Built environment features, such as housing structure and road density depict the physical context affecting mobility, accessibility and exposure to hazard, which further affect evacuation rate. Similarly, evacuation readiness indicators, such as vehicle access ratio reflect the logistical conditions that may affect a community's ability to respond quickly. These factors can compound the effects of community vulnerability and hazard intensity to shape evacuation behaviors.

While the ranking of the dimensions is the same, their relative magnitudes reveal important differences shaped by the nature of each hazard. For hurricanes, the hazard intensity dimension is overwhelmingly dominant, with an aggregated importance value that is significantly more influential relative to all other categories. This signifies that for large-scale, slow-onset events with ample warning time and widespread media coverage, the decision to evacuate is heavily rationalized based on a few clear, officially communicated metrics like storm intensity and evacuation zone mandates. In contrast, for wildfires, the gap between the importance of hazard intensity and community vulnerability is significantly smaller. This indicates that for rapid-onset, localized, and more uncertain hazards, the decision-making process is more balanced. While the effect of immediate threat is important, residents' evacuation behaviors are also characterized by other factors, such as socio-economic capacity. This suggests that when official information is rapidly changing or less certain, residents may rely more heavily on their current resources and vulnerabilities to guide their decision to evacuate.

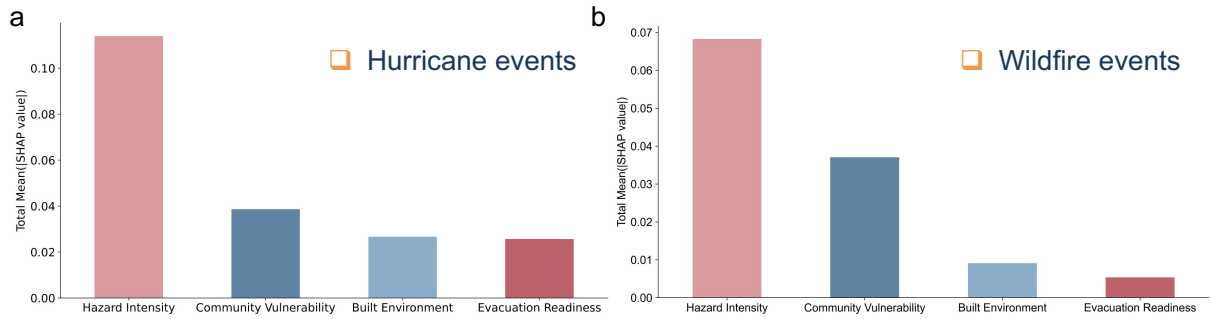

**Fig.6** Feature dimension contributions for a) hurricane events, b) wildfire events. Both plots show a consistent ranking across the four feature dimensions, with hazard intensity ranked highest, followed by community vulnerability, built environment and evacuation readiness.

## 5 Concluding remarks

This paper introduces a novel framework to predict hazard evacuation rates and to understand their underlying drivers using an interpretable machine learning approach. By integrating multi-source data across four key dimensions, hazard intensity, community vulnerability, evacuation readiness, and the built environment, this study developed and evaluated hazard-specific models for major hurricane and wildfire events in the United States. This study yields several key findings that hold significant academic contributions and practical implications for disaster management.

First, the interpretable machine learning model presented in this study, Evac-cast, provides an effective and powerful tool for predicting evacuation behavior. The high predictive accuracy of the XGBoost models confirms their ability to capture the complex, non-linear relationships between the multifaceted drivers of evacuation. This work provides a robust methodological alternative to traditional linear models, showcasing how non-linear approaches can better handle the complexity inherent in human behavioral responses during disasters. Emergency management agencies can leverage this approach to develop more accurate, real-time evacuation demand forecasting tools.

Using readily available data, these models can provide agencies with timely and granular predictions, improving resource allocation, traffic management, and overall operational planning in the critical hours before a disaster strikes.

Second, through the use of SHAP, this study moved beyond black-box predictions to identify a clear hierarchy of factors influencing evacuation decisions. Understanding which factors are most influential enables more targeted and effective interventions. Specifically, the analysis showed that for hurricanes, evacuation is driven primarily by clear indicators of threat, such as max wind speed and official evacuation zone locations. The finding confirms that official public messaging should continue to prioritize these clear, potent metrics. Similarly, the most critical factors for wildfire are the immediate fire proximity and fire hazard severity zone, followed by socio-economic indicators, such as high school educational attainment and weather conditions. For wildfires, the dominance of fire proximity highlights the need for real-time, neighborhood-level threat mapping and communication. Furthermore, the importance of education level in wildfire contexts suggests that communication strategies must be designed to be easily and quickly understood by all populations, especially during fast-moving, uncertain events. This research answers the call for more explainable AI in disaster science by demonstrating how SHAP values can be combined with XGBoost models to move beyond black-box predictions of evacuation rates. Specifically, it provides a method not only to forecast granular evacuation outcomes but also to precisely quantify the impact of context-dependent drivers, thereby bridging the gap between predictive power and a nuanced theoretical understanding of evacuation behavior.

Third, the cross-hazard comparison revealed a consistent, universal pattern in the hierarchy of evacuation drivers. Across both hurricanes and wildfires, hazard intensity was found to be the most critical dimension, followed in order by community vulnerability, built environment, and evacuation readiness. This finding provides strong empirical evidence for a universal, structured decision-making framework in hazard response, contributing a new, generalizable framework to the literature that has not been fully explored yet. The existence of a universal pattern of evacuation determinants confirms that the primary task for emergency managers, regardless of the threat, is the clear and effective communication of the hazard's intensity. This allows for the development of more streamlined and robust public warning systems that can be adapted to various disaster scenarios.

Finally, the good performance of the model in predicting evacuation patterns without involving explicit psychosocial attributes challenges the conventionally emphasized role of these factors in evacuation prediction. For decades, evacuation research has extensively studied the role of psychosocial factors, including risk perception and social ties, assuming their necessity for understanding evacuation. While the importance of those factors like risk perception and social ties is well-established, these attributes are often difficult to acquire in a timely and large-scale manner, limiting their use in pre-hazard operational models. Our findings show that it is possible to achieve strong predictive performance without relying on this type of data. This outcome does not diminish the theoretical importance of psychosocial factors but rather provides a new perspective on their application. It suggests that the collective influence of these micro-level behavioral drivers may be effectively captured by more accessible, macro-level proxy variables (e.g., socio-demographics and hazard intensity). This contributes to the literature by demonstrating a viable approach for large-scale behavioral modeling where individual-level data is unavailable and opens new research questions into the relationship between these different scales of analysis. This finding has significant implications for operational forecasting. It demonstrates that emergency management agencies do not need to be constrained by the lack of costly and time-consuming survey data to develop effective predictive tools. By leveraging existing, publicly available datasets, agencies can build powerful models to enable scalable, real-time evacuation demand forecasting. This advantage ultimately provides a more agile and data-driven approach to support life-saving decisions in the critical window before a disaster.

This study's limitations highlight important avenues for future research. First, while this study incorporated several features of the built environment, such as road and building density, it did not explicitly consider the topological properties of the transportation network. Factors like road connectivity and network centrality are important determinants of the actual feasibility and perceived ease of evacuation. Future work could integrate these network-based metrics to create a more sophisticated representation of the built environment, which may improve model accuracy and provide more granular insights for traffic management and route planning. Second, the cross-hazard analysis, while a key strength, was limited to hurricanes and wildfires. To further test the universality of the identified driver hierarchy, future research should expand this framework to other hazard scenarios. Applying the model to different types of natural hazards, such as earthquakes or floods, in various geographic and cultural contexts would provide a more rigorous test of the universal patterns observed.

**Acknowledgement**

This work was supported by the National Science Foundation under Grant CMMI-1846069 (CAREER). The authors also would like to acknowledge the data support from Cuebiq. Any opinions, findings, conclusions, or recommendations expressed in this research are those of the authors and do not necessarily reflect the views of the National Science Foundation or Cuebiq.

# References

Anyidoho, P. K., Davidson, R. A., Rambha, T., & Nozick, L. K. (2022). Prediction of population behavior in hurricane evacuations. Transportation research part A: policy and practice, 159, 200-221.

Ao, Y., Huang, K., Wang, Y., Wang, Q., & Martek, I. (2020). Influence of built environment and risk perception on seismic evacuation behavior: Evidence from rural areas affected by Wenchuan earthquake. International Journal of Disaster Risk Reduction, 46, 101504.

Bakhshian, E., & Martinez-Pastor, B. (2023). Evaluating human behaviour during a disaster evacuation process: A literature review. Journal of traffic and transportation engineering (English edition), 10(4), 485-507.

Balboa, A., Cuesta, A., González-Villa, J., Ortiz, G., & Alvear, D. (2024). Logistic regression vs machine learning to predict evacuation decisions in fire alarm situations. Safety Science, 174, 106485.

Bing Maps. GlobalMLBuildingFootprints. Retrieved from: https://github.com/microsoft/GlobalMLBuildingFootprints?tab=readme-ov-file

Branco, P., Torgo, L., & Ribeiro, R. P. (2017). SMOGN: a pre-processing approach for imbalanced regression. Paper presented at the First international workshop on learning with imbalanced domains: Theory and applications.

Brooks, B. (2022). Millions in Florida urged to evacuate as Hurricane Ian nears. Retrieved from https://www.reuters.com/world/us/hurricane-ian-nears-millions-florida-told-evacuate-2022-09-27/

Buylova, A., Chen, C., Cramer, L. A., Wang, H., & Cox, D. T. (2020). Household risk perceptions and evacuation intentions in earthquake and tsunami in a Cascadia Subduction Zone. International Journal of Disaster Risk Reduction, 44, 101442.

CALFIRE. California Historical Fire Perimeters. Retrieved from https://data.ca.gov/dataset/california-historical-fire-perimeters/resource/721d443a-200f-4943-b886-6af2f6a18b99

CALFIRE. (2024). Fire Hazard Severity Zones, in SRA Effective April 1, 2024 with LRA Recommended 2007-2011. Retrieved from https://data.ca.gov/dataset/fire-hazard-severity-zones-in-sra-effective-april-1-2024-with-lra-recommended-2007-2011

Charlotte Phillipp. (2025). Eaton and Palisades Wildfires Are Fully Contained. Retrieved from https://people.com/eaton-and-palisades-wildfires-are-fully-contained-authorities-8780189?

Chen, T., & Guestrin, C. (2016). Xgboost: A scalable tree boosting system. Paper presented at the Proceedings of the 22nd acm sigkdd international conference on knowledge discovery and data mining.

Collins, J., Polen, A., McSweeney, K., Colón-Burgos, D., & Jernigan, I. (2021). Hurricane risk perceptions and evacuation decision-making in the age of COVID-19. Bulletin of the American Meteorological Society, 102(4), E836-E848.

County of Sonoma. (2020). 2019 KINCADE FIRE AFTER ACTION REPORT. Retrieved from https://sonomacounty.ca.gov/Main%20County%20Site/Administrative%20Support%20%26%20Fiscal%20Services/Emergency%20Management/Documents/Archive/Administration/Services/Training__3/Service%201/_Documents/Sonoma-County-2019-Kincade-Fire-AAR-FINAL-ADA.pdf?

Dash, N., & Gladwin, H. (2007). Evacuation decision making and behavioral responses: Individual and household. Natural Hazards Review, 8(3), 69-77.

Demuth, J. L., Morss, R. E., Lazo, J. K., & Trumbo, C. (2016). The effects of past hurricane experiences on evacuation intentions through risk perception and efficacy beliefs: A mediation analysis. Weather, Climate, and Society, 8(4), 327-344.

Ebrahimi, H., Sattari, F., Lefsrud, L., & Macciotta, R. (2023). A machine learning and data analytics approach for predicting evacuation and identifying contributing factors during hazardous materials incidents on railways. Safety Science, 164, 106180.

Florida Climate Center. Hurricanes. Retrieved from https://climatecenter.fsu.edu/topics/hurricanes

Florida Division of Emergency Management. Know Your Zone, Know Your Home. Retrieved from https://www.floridadisaster.org/knowyourzone/

Forrister, A., Kuligowski, E. D., Sun, Y., Yan, X., Lovreglio, R., Cova, T. J., & Zhao, X. (2024). Analyzing risk perception, evacuation decision and delay time: a case study of the 2021 Marshall Fire in Colorado. Travel behaviour and society, 35, 100729.

Fusco, G., Zhu, J., & Yang, J. (2025). Investigating the Impacts of Built Environment Knowledge on Hurricane Evacuation Intentions in a College Student Sample: A Test of the Power of Information. Natural Hazards Review, 26(1), 04024054.

Golshani, N., Shabanpour, R., Mohammadian, A., Auld, J., & Ley, H. (2019). Evacuation decision behavior for no-notice emergency events. Transportation Research Part D: Transport and Environment, 77, 364-377.

Guan, X., & Chen, C. (2021). A behaviorally-integrated individual-level state-transition model that can predict rapid changes in evacuation demand days earlier. Transportation research part E: logistics and transportation review, 152, 102381.

Halim, N., Jiang, F., Meng, S., Mozumder, P., & Yao, C. (2024). Traveling for Safety: Price and Income Elasticities of Hurricane Evacuation Behavior. Transportation Research Record, 03611981241292593.

Hasan, S., Mesa-Arango, R., Ukkusuri, S., & Murray-Tuite, P. (2012). Transferability of hurricane evacuation choice model: Joint model estimation combining multiple data sources. Journal of Transportation Engineering, 138(5), 548-556.

Huang, S.-K., Lindell, M. K., & Prater, C. S. (2016). Who leaves and who stays? A review and statistical meta-analysis of hurricane evacuation studies. Environment and Behavior, 48(8), 991-1029.

Huang, S.-K., Lindell, M. K., & Prater, C. S. (2017). Multistage model of hurricane evacuation decision: Empirical study of Hurricanes Katrina and Rita. Natural Hazards Review, 18(3), 05016008.

Huang, S.-K., Lindell, M. K., Prater, C. S., Wu, H.-C., & Siebeneck, L. K. (2012). Household evacuation decision making in response to Hurricane Ike. Natural Hazards Review, 13(4), 283-296.

Julia Reinstein. (2024). Hurricane Milton: Millions under evacuation orders as storm closes in on Florida. Retrieved from https://abcnews.go.com/US/florida-officials-give-evacuation-orders-hurricane-milton-strengthens/story?id=114562685

Kuligowski, E. D., Zhao, X., Lovreglio, R., Xu, N., Yang, K., Westbury, A., . . . Brown, N. (2022). Modeling evacuation decisions in the 2019 Kincade fire in California. Safety Science, 146, 105541.

Lee, C.-C., Chou, C., & Mostafavi, A. (2022). Specifying evacuation return and home-switch stability during short-term disaster recovery using location-based data. Scientific Reports, 12(1), 15987.

León, J., Gubler, A., & Ogueda, A. (2022). Modelling geographical and built-environment attributes as predictors of human vulnerability during tsunami evacuations: a multi-case-study and paths to improvement. Natural Hazards and Earth System Sciences, 22(9), 2857-2878.

Li, B., & Mostafavi, A. (2022). Location intelligence reveals the extent, timing, and spatial variation of hurricane preparedness. Scientific Reports, 12(1), 16121. doi:10.1038/s41598-022-20571-3

Li, J., & Ozbay, K. (2015). Hurricane Irene evacuation traffic patterns in New Jersey. Natural Hazards Review, 16(2), 05014006.

Lindell, M. K., Lu, J.-C., & Prater, C. S. (2005). Household decision making and evacuation in response to Hurricane Lili. Natural Hazards Review, 6(4), 171-179.

Lindell, M. K., Murray-Tuite, P., Wolshon, B., & Baker, E. J. (2018). Large-scale evacuation: The analysis, modeling, and management of emergency relocation from hazardous areas: CRC Press.

Lindell, M. K., & Perry, R. W. (1992). Behavioral foundations of community emergency planning: Hemisphere Publishing Corp.

Lindell, M. K., & Perry, R. W. (2003). Communicating environmental risk in multiethnic communities: Sage publications.

Liu, L., Zhang, X., Jiang, S., & Zhao, X. (2025). Hurricane evacuation analysis with large-scale mobile device location data during hurricane Ian. Transportation Research Part D: Transport and Environment, 139, 104559.

Lovreglio, R., Ronchi, E., & Nilsson, D. (2016). An Evacuation Decision Model based on perceived risk, social influence and behavioural uncertainty. Simulation Modelling Practice and Theory, 66, 226-242.

Lundberg, S. M., & Lee, S.-I. (2017). A unified approach to interpreting model predictions. Advances in neural information processing systems, 30.

Manzhu Yu. (2025). Q&A: Causes, spread and solutions for California's wildfire crisis. Retrieved from https://www.psu.edu/news/research/story/qa-causes-spread-and-solutions-californias-wildfire-crisis?

Metaxa-Kakavouli, D., Maas, P., & Aldrich, D. P. (2018). How social ties influence hurricane evacuation behavior. Proceedings of the ACM on human-computer interaction, 2(CSCW), 1-16.

Morss, R. E., Cuite, C. L., & Demuth, J. L. (2024). What predicts hurricane evacuation decisions? The importance of efficacy beliefs, risk perceptions, and other factors. npj Natural Hazards, 1(1), 24.

Mozumder, P., & Vásquez, W. F. (2018). Understanding hurricane evacuation decisions under contingent scenarios: A stated preference approach. Environmental and resource economics, 71, 407-425.

Murray-Tuite, P., Yin, W., Ukkusuri, S. V., & Gladwin, H. (2012). Changes in evacuation decisions between Hurricanes Ivan and Katrina. Transportation Research Record, 2312(1), 98-107.

National Centers for Environmental Information. (2025). Billon-Dollar Weather and Climate Disasters. Retrieved from https://www.ncei.noaa.gov/access/billions/

National Hurricane Center. (2023). Hurricane Ian. Retrieved from https://www.nhc.noaa.gov/data/tcr/AL092022_Ian.pdf

National Oceanic And Atmospheric Administration. NHC GIS Archive - Tropical Cyclone Best Track.

National Weather Service. (2024). Hurricane Milton Impacts to East Central Florida. Retrieved from https://www.weather.gov/mlb/HurricaneMilton_Impacts?

Open-Meteo. Historical Forecast API. Retrieved from: https://open-meteo.com/en/docs/historical-forecast-api

Rambha, T., Nozick, L. K., & Davidson, R. (2021). Modeling hurricane evacuation behavior using a dynamic discrete choice framework. Transportation research part B: methodological, 150, 75-100.

Rashid, M. M., Tirtha, S. D., Eluru, N., & Hasan, S. (2025). Understanding hurricane evacuation behavior from Facebook data. International Journal of Disaster Risk Reduction, 116, 105147.

Riad, J. K., Norris, F. H., & Ruback, R. B. (1999). Predicting evacuation in two major disasters: Risk perception, social influence, and access to resources 1. Journal of applied social Psychology, 29(5), 918-934.

Roy, K. C., Hasan, S., Abdul-Aziz, O. I., & Mozumder, P. (2022). Understanding the influence of multiple information sources on risk perception dynamics and evacuation decisions: An agent-based modeling approach. International Journal of Disaster Risk Reduction, 82, 103328.

Sadri, A. M., Ukkusuri, S. V., & Gladwin, H. (2017). The role of social networks and information sources on hurricane evacuation decision making. Natural Hazards Review, 18(3), 04017005.

Sun, Y., Forrister, A., Kuligowski, E. D., Lovreglio, R., Cova, T. J., & Zhao, X. (2024). Social vulnerabilities and wildfire evacuations: A case study of the 2019 Kincade fire. Safety Science, 176, 106557.

Sun, Y., Huang, S.-K., & Zhao, X. (2024). Predicting hurricane evacuation decisions with interpretable machine learning methods. International Journal of Disaster Risk Science, 15(1), 134-148.

Tanim, S. H., Wiernik, B. M., Reader, S., & Hu, Y. (2022). Predictors of hurricane evacuation decisions: A meta-analysis. Journal of Environmental Psychology, 79, 101742.

Thakur, S., Ranjitkar, P., & Rashidi, S. (2022a). Investigating evacuation behaviour under an imminent threat of volcanic eruption using a logistic regression-based approach. Safety Science, 149, 105688.

Thakur, S., Ranjitkar, P., & Rashidi, S. (2022b). Modelling evacuation decisions under a threat of volcanic eruption in Auckland. Transportation Research Part D: Transport and Environment, 109, 103374.

Thompson, R. R., Garfin, D. R., & Silver, R. C. (2017). Evacuation from natural disasters: a systematic review of the literature. Risk Analysis, 37(4), 812-839.

U.S. Census Bureau. 2024 TIGER/Line Shapefiles: Roads. Retrieved from: https://www.census.gov/cgi-bin/geo/shapefiles/index.php?year=2024&layergroup=Roads

U.S. Census Bureau. American Community Survey Retrieved from https://www.census.gov/data.html

U.S. Census Bureau. Tiger/Line Shapefiles. Retrieved from https://www.census.gov/geographies/mapping-files/time-series/geo/tiger-line-file.html

Washington, V., Guikema, S., Mondisa, J. L., & Misra, A. (2024). A data-driven method for identifying the locations of hurricane evacuations from mobile phone location data. Risk Analysis, 44(2), 390-407.
Wong, S. D., Pel, A. J., Shaheen, S. A., & Chorus, C. G. (2020). Fleeing from hurricane Irma: Empirical analysis of evacuation behavior using discrete choice theory. Transportation Research Part D: Transport and Environment, 79, 102227.
Woo, M., Hui, K. T. Y., Ren, K., Gan, K. E., & Kim, A. (2017). Reconstructing an emergency evacuation by ground and air the wildfire in Fort McMurray, Alberta, Canada. Transportation Research Record, 2604(1), 63-70.
Wu, A., Yan, X., Kuligowski, E., Lovreglio, R., Nilsson, D., Cova, T. J., . . . Zhao, X. (2022). Wildfire evacuation decision modeling using GPS data. International Journal of Disaster Risk Reduction, 83, 103373.
Xu, N., Lovreglio, R., Kuligowski, E. D., Cova, T. J., Nilsson, D., & Zhao, X. (2023). Predicting and assessing wildfire evacuation decision-making using machine learning: Findings from the 2019 kincade fire. Fire Technology, 59(2), 793-825.
Yabe, T., Sekimoto, Y., Tsubouchi, K., & Ikemoto, S. (2019). Cross-comparative analysis of evacuation behavior after earthquakes using mobile phone data. Plos one, 14(2), e0211375.
Younes, H., Darzi, A., & Zhang, L. (2021). How effective are evacuation orders? An analysis of decision making among vulnerable populations in Florida during hurricane Irma. Travel behaviour and society, 25, 144-152.
Zhang, R., Liu, D., Xu, Y., Xu, C., & Chen, X. (2024). Personal factors influencing emergency evacuation decisions under different flash flood characteristics. Natural Hazards, 1-27.
Zhang, X. A., & Borden, J. (2024). A Cross-Community Comparison of Antecedents of Hurricane Ian Risk Perceptions and Evacuation Behaviours. Journal of Contingencies and Crisis Management, 32(4), e70009.
Zhao, X., Lovreglio, R., & Nilsson, D. (2020). Modelling and interpreting pre-evacuation decision-making using machine learning. Automation in Construction, 113, 103140.
40